\documentclass[letter]{aa} % for the letters 
\usepackage{graphicx}
\usepackage{txfonts}
\usepackage{natbib}
\bibpunct{(}{)}{;}{a}{}{,}

\begin{document}
   \title{Gamma-Ray Burst afterglows : luminosity clustering at infrared wavelengths ?}

   \author{B. Gendre
          \inst{1}
          \and
          S. Pelisson\inst{1}
\and	M. Bo\"er\inst{2}
	\and
	S. Basa\inst{1}
	\and
	A. Mazure\inst{1}
          }

   \institute{Laboratoire d'Astrophysique de Marseille/CNRS/Universite de Provence, Technopole de l'Etoile, 38 rue Frederic Joliot-Curie, 13388 Marseille cedex 13, FRANCE
         \and
	Observatoire de Haute Provence/CNRS, 04870 St.Michel l'Observatoire, FRANCE
             }

   \date{}

% \abstract{}{}{}{}{} 
% 5 {} token are mandatory
 
  \abstract
   { Clustering in the luminosity of the afterglows of gamma-ray burst has been reported in the optical and X-ray.}
   { We investigate the possibility that a clustering in the luminosity of the afterglows of gamma-ray burst exists in the near infrared (J, H, K bands). We use observations of events from 1997 to the end of 2007.}
   {We correct the gamma-ray burst afterglow light curve for distance effects and time dilation, and rescale all light curves to a common distance of $z=1$. We used only observations of signals emitted in the near infrared (in the burst frame).}
   {We observe a clustering identical to the one observed in the optical and similar to the one observed in X-rays. We thus confirm previous work made in optical wavelengths and set a constraint on the total energy of the fireball.}
   {}


   \maketitle
%
%________________________________________________________________

\section{Introduction}

Long Gamma-Ray Bursts (GRBs) are located at cosmological distances \citep{met97}. These events are associated with supernovae \citep[e.g.][]{hjo03, sta03} and derive from the endpoint of the stellar evolution of massive stars. It is thus tempting to hypothesize a standard candle behavior as for type Ia-supernovae. Indeed, several empirical correlations between observable quantities have been discovered \citep{ama02, ghi03}, triggering the search for a standard candle based on GRBs \citep[see e.g.][]{ghi06}. This could have a very high impact on studies of the cosmology in the redshift range 1-15 \citep[for a review, see][]{mes06}. However, the use of GRBs for such studies requires the establishment of a robust indicator of their distance based on their intrinsic properties. This, in turn, implies that we must fully understand the properties of GRBs and their luminosity function.

During the BeppoSAX era, we discovered a clustering of the X-ray luminosities of GRB afterglows \citep{boe00}. This clustering was confirmed later \citep{gen05}, and was also observed in the optical \citep{nar06, kan06, lia06}. To date, GRBs cluster in three different groups according to their optical and X-ray luminosities; however, 10 \% of nearby outliers are apparently underluminous and do not cluster \citep{gen08}. Because of selection effects, this population of underluminous afterglows could be the tip of the iceberg of a more numerous undetected population.

The physical reason for the clustering in luminosity is still unknown. It has been proposed that the microphysical parameters of the fireball (which fix the afterglow properties) can take on only a few possible values \citep{gen08}. To confirm this hypothesis, we must investigate the luminosity clustering at longer wavelengths (compared to optical bands), so that the fireball parameters can be constrained.  In this Letter, we present our findings obtained using infrared observations. They are interesting because :

\begin{itemize}
\item infrared observations are numerous and can be used as a statistically significant sample to estimate the positions of characteristic breaks within the afterglow broad-band spectrum, such as the injection frequency $\nu_m$;
\item the infrared sky is less sensitive to extinction compared to optical wavelengths \citep{pei92}. Because the extinction law (and thus the extinction correction) in distant galaxies is not well known \citep{str04}, this lowers the flux uncertainties due to the extinction correction.
\end{itemize}

While our sample is not large enough to draw definitive conclusions, in section \ref{sec_data} we show that there is evidence of luminosity clustering in the infrared. We will discuss this result in the fireball model framework in section \ref{sec_discu}.

%__________________________________________________________________

\section{Data reduction and analysis}
\label{sec_data}

\subsection{Sample selection and definition}

\begin{table}
\begin{minipage}[t]{\columnwidth}
\caption{Sample of GRB afterglows used. We indicate for each burst its redshift, the bands observed at Earth and the references for the light curves used.}
\label{table_sample}
\centering
\renewcommand{\footnoterule}{}  % to avoid a line before footnotes
\begin{tabular}{cccc}
\hline \hline
GRB                  & Redshift &  Observation band    & Reference  \\
%    &(J2000) & (J2000) &~    & [{\rm mJy}] \\
\hline
\object{GRB 970228}  & 0.70 & H,K & \footnote{\citet{gal97}, \citet{fru99}, \citet{gal00}} \\
\object{GRB 980703}  & 0.97 & H,K & \footnote {\citet{vre99}, \citet{cas01}, \citet{blo98}} \\
\object{GRB 990123}  & 1.60 &  K' & \footnote{\citet{kul99}, \citet{hol04}} \\
\object{GRB 991208}  & 0.71 &  K  & \footnote{\citet{sag00}, \citet{cas01}} \\
\object{GRB 991216}  & 1.02 & H,K & \footnote{\citet{hal00}, \citet{ber01}, \citet{sag00}} \\
\object{GRB 000418}  & 1.12 & K,Ks,K'& \footnote{\citet{kls00}, \citet{gor03}} \\
\object{GRB 000911}  & 1.06 & H,Ks & \footnote{\citet{laz01}, \citet{mas05}} \\
\object{GRB 010921}  & 0.45 & J,H,Ks & \footnote{\citet{pri02}} \\
\object{GRB 011121}  & 0.36 & J,H,K,Ks & \footnote{\citet{prc02}, \citet{gre03}} \\
\object{GRB 020405}  & 0.69 & H,K & \footnote{\citet{pri03}, \citet{str05}} \\
\object{GRB 021211}  & 1.01 & H,K & \footnote{\citet{hol04}} \\
\object{GRB 030329}  & 0.17 & J,H &\footnote{\citet{blo04}} \\
\object{GRB 030528}  & 0.78 & H,K,Ks & \footnote{\citet{rau04}, \citet{rau05}} \\
\object{GRB 031203}  & 0.11 & J,H,K &\footnote{\citet{gal04}, \citet{mal04}} \\
%\object{GRB 041006}  & 0.72 &  H  & \footnote{\citet{sod06}} \\
\object{GRB 050126}  & 1.29 &  Ks & \footnote{\citet{ber05}} \\
\object{GRB 050408}  & 1.24 &  K  & \footnote{\citet{fol06}} \\
%\object{GRB 050416A} & 0.65 &  Ks & \footnote{\citet{hol07}} \\
%\object{GRB 051111}  & 1.55 &  Ks & \footnote{\citet{but06}} \\
%\object{GRB 060418}  & 1.49 &  K' & \footnote{\citet{mol07}} \\
\object{GRB 060614}  & 0.125 &  J  & \footnote{\citet{del06}, \citet{cob06}} \\
%\object{GRB 061126}  & 1.16 &  Ks & \footnote{\citet{per08}} \\
\object{GRB 070125}  & 1.55 &  Ks  & \footnote{\citet{cha08}} \\
\object{GRB 070306}  & 1.50 &  K   & \footnote{\citet{jau08}} \\
\object{GRB 071010A} & 0.98 &  H,K & \footnote{\citet{cov08}} \\
%\object{GRB 000000}  &  &  & \\
%\object{GRB 000000}  &  &  & \\
%\object{GRB 000000}  &  &  & \\
%\object{GRB 000000}  &  &  & \\
%\object{GRB 000000}  &  &  & \\
%\object{GRB 000000}  &  &  & \\
%\object{GRB 000000}  &  &  & \\
%\object{GRB 000000}  &  &  & \\
%\object{GRB 000000}  &  &  & \\
%\object{GRB 000000}  &  &  & \\
%\object{GRB 000000}  &  &  & \\
\hline
\end{tabular}
\end{minipage}
\end{table}

   \begin{figure*}
   \centering
   \includegraphics[width=6.2cm]{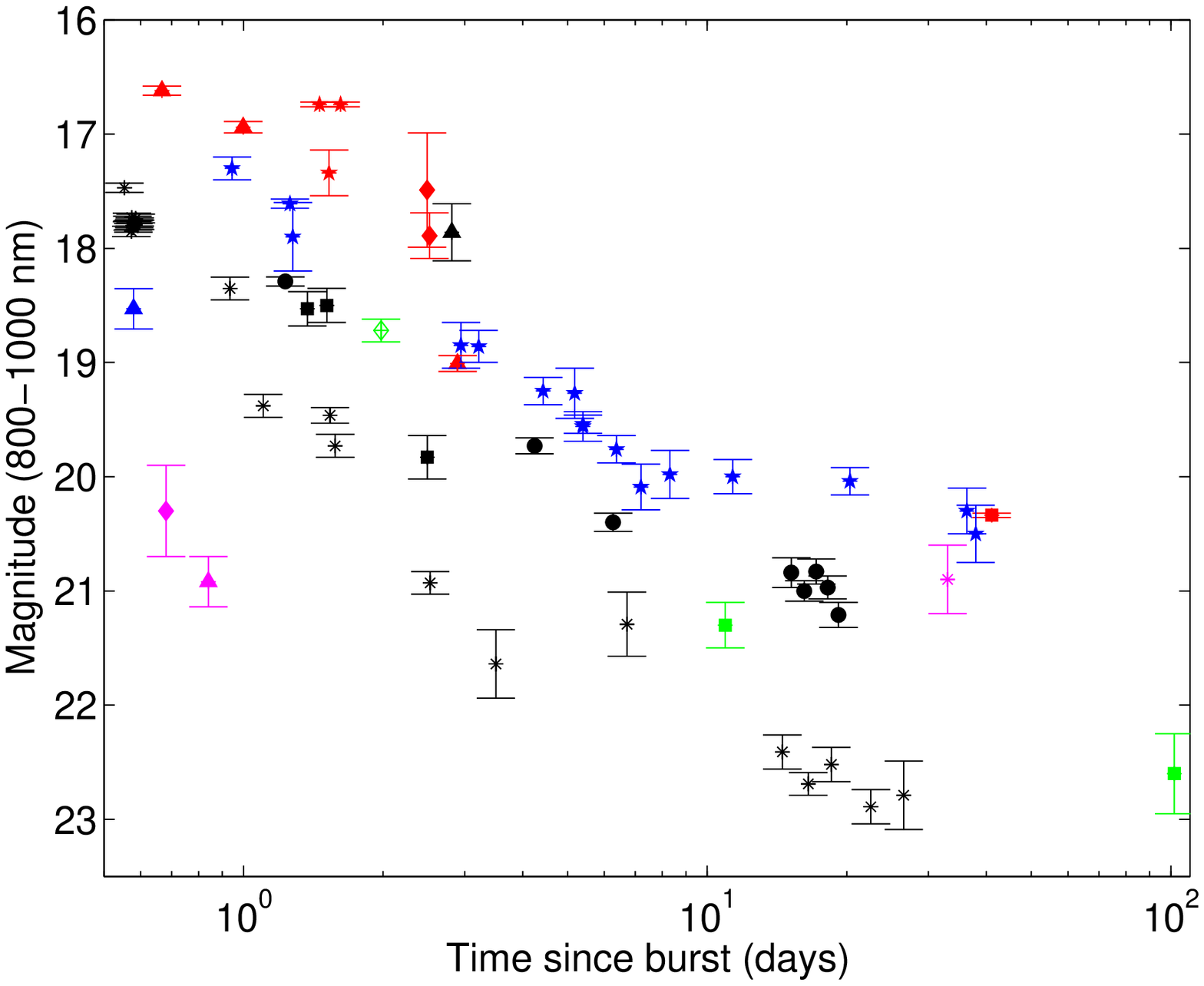}
   \includegraphics[width=6.2cm]{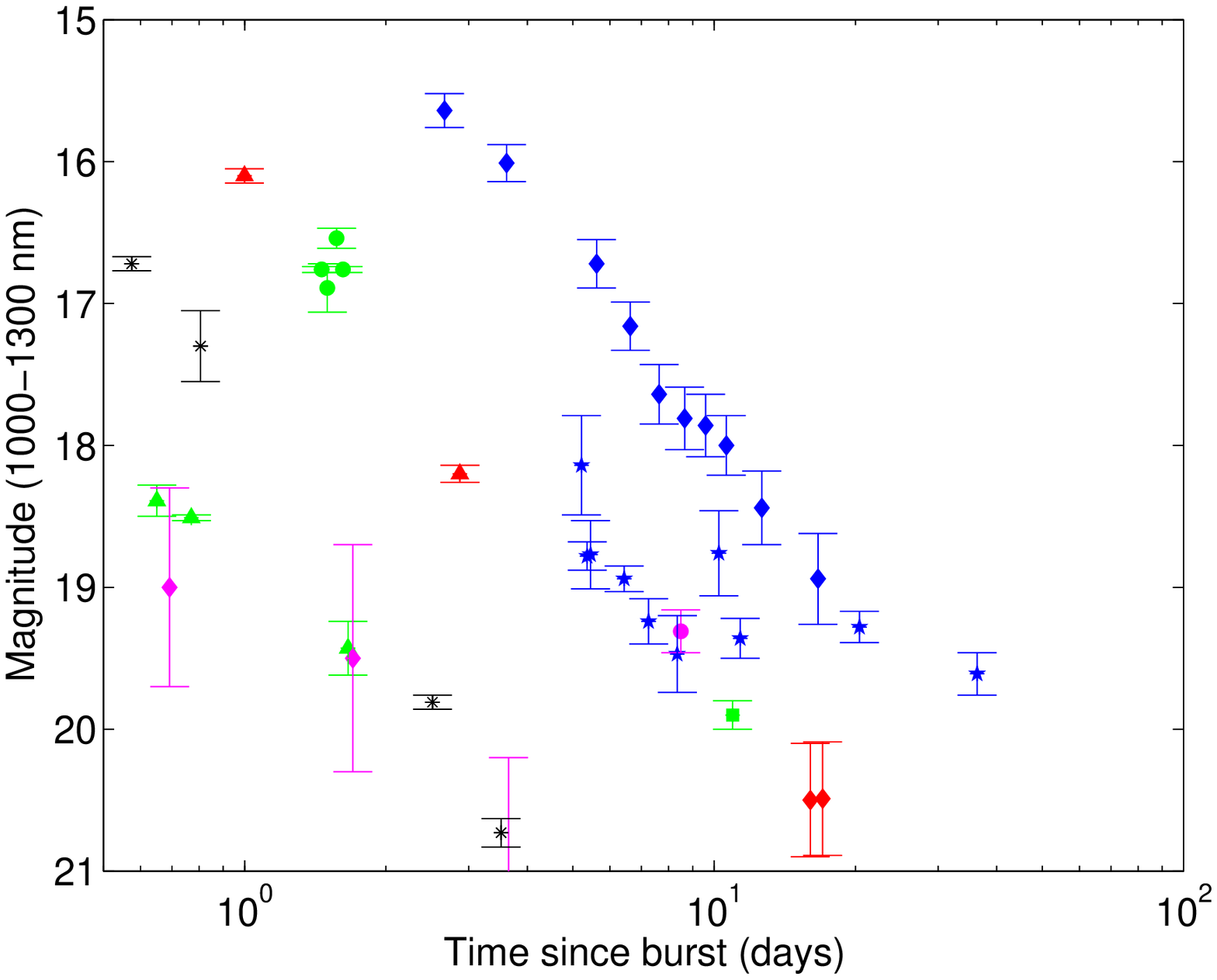}
   \includegraphics[width=5.4cm]{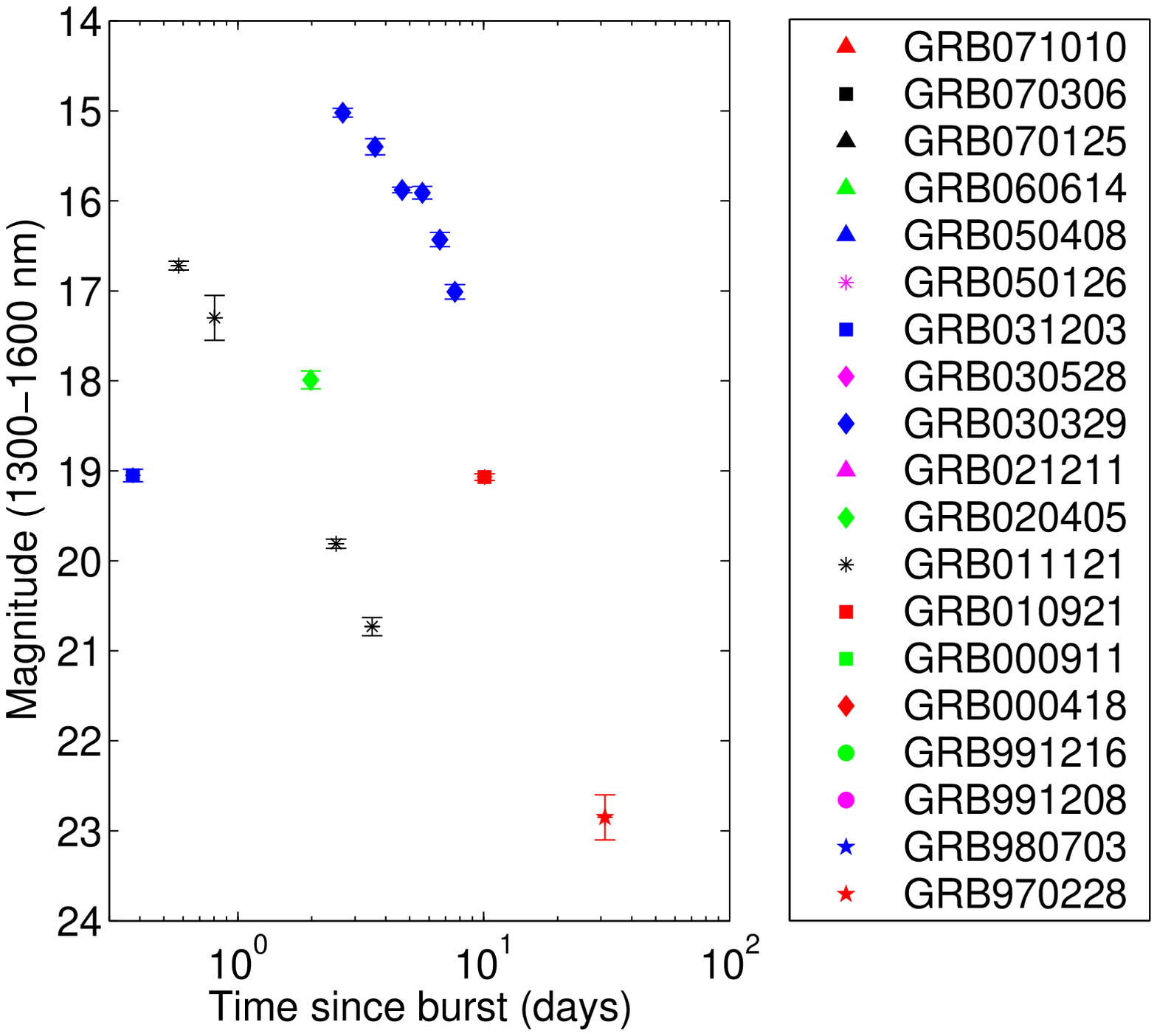}
      \caption{{\bf Light curves of our} sample of afterglows, without distance correction. We plot them using the three bands we chose (see text for details) : the $8000-10000$~\AA~band (left), the $10000-13000$~\AA~band (middle), the $13000-16000$~\AA~band (right). Each burst is represented by a combination of different symbols and colors. See the electronic version for colors.
              }
         \label{Fig_raw}
   \end{figure*}

  \begin{figure*}
   \centering
   \includegraphics[width=6cm]{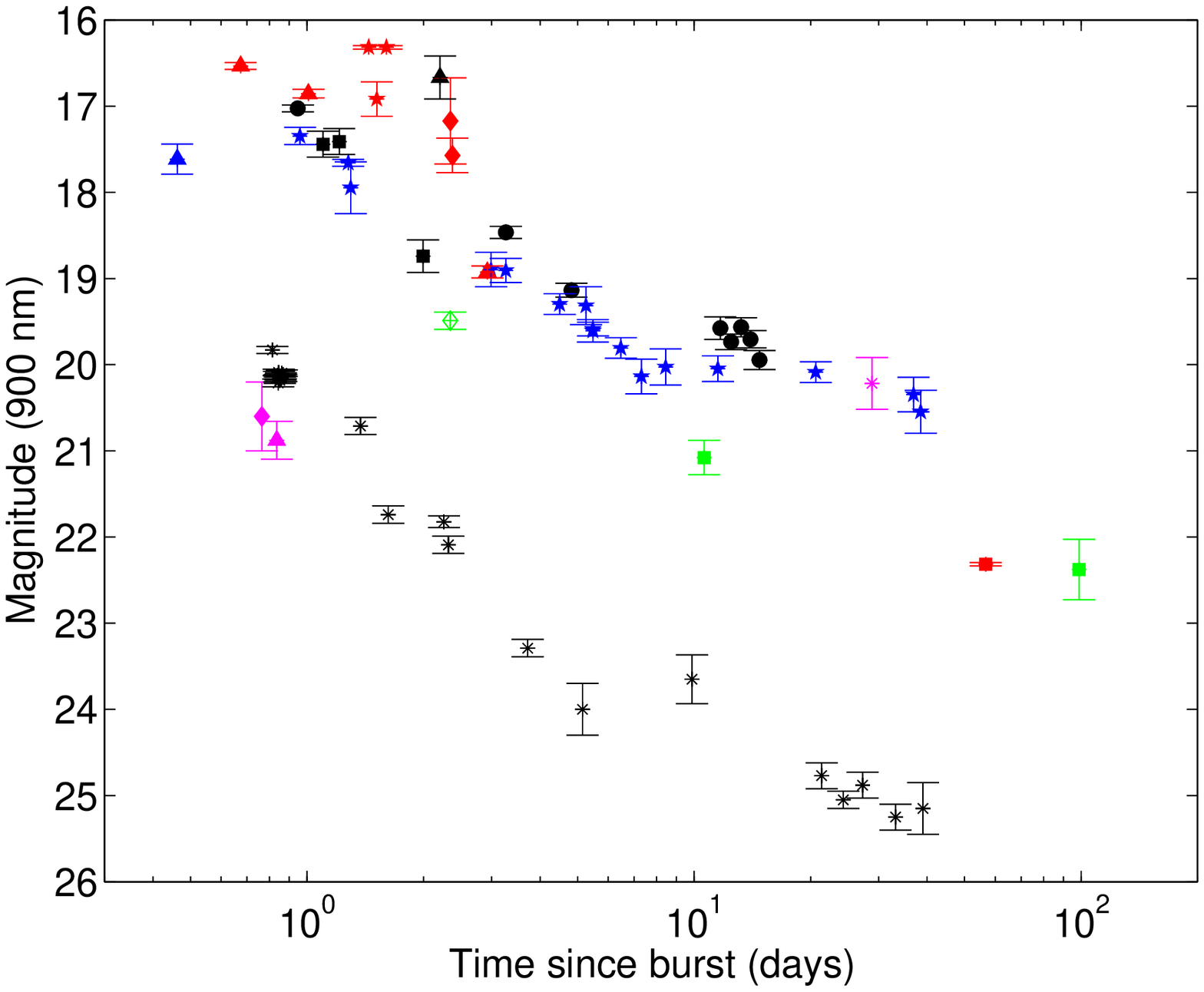}
   \includegraphics[width=6cm]{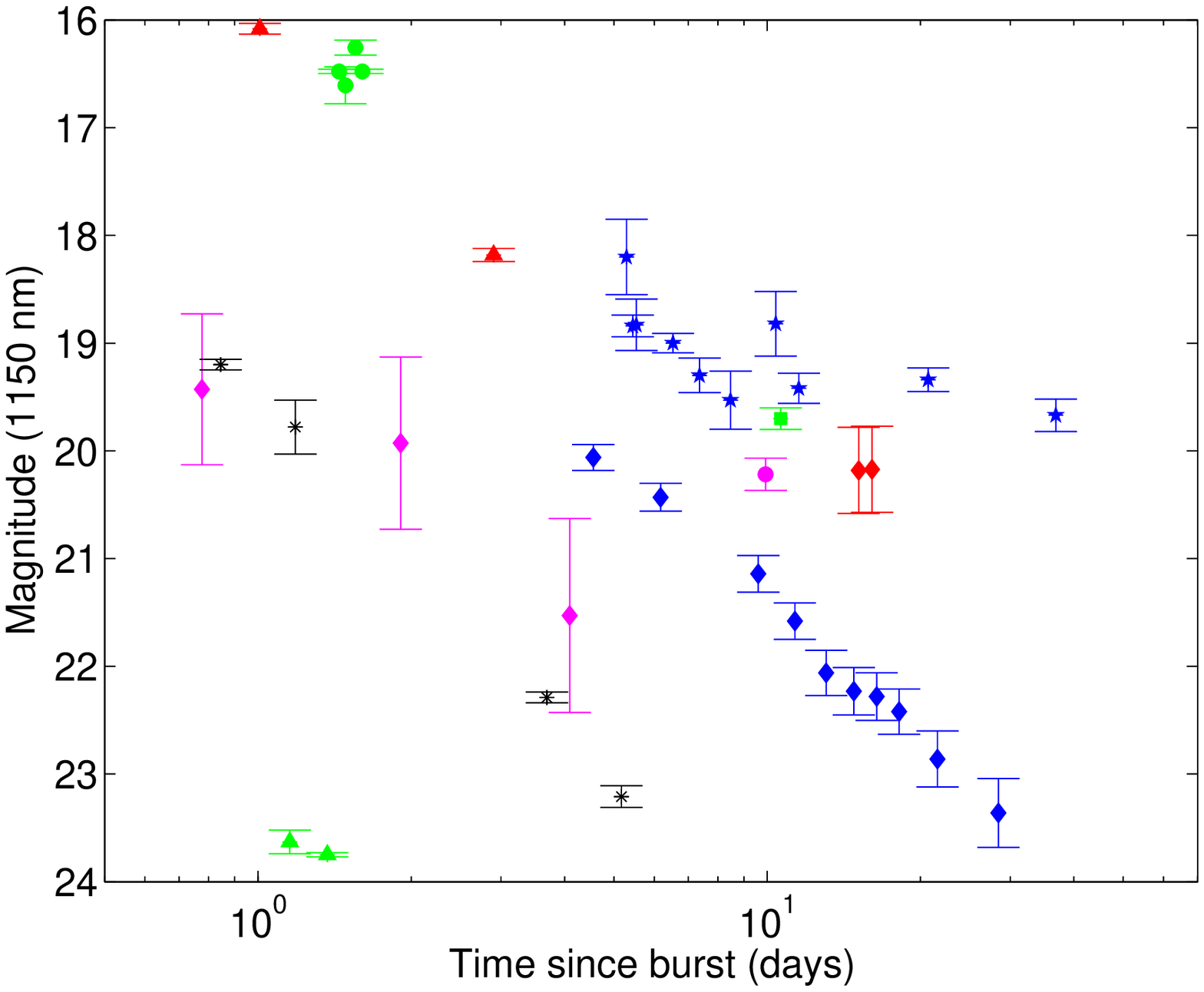}
   \includegraphics[width=6cm]{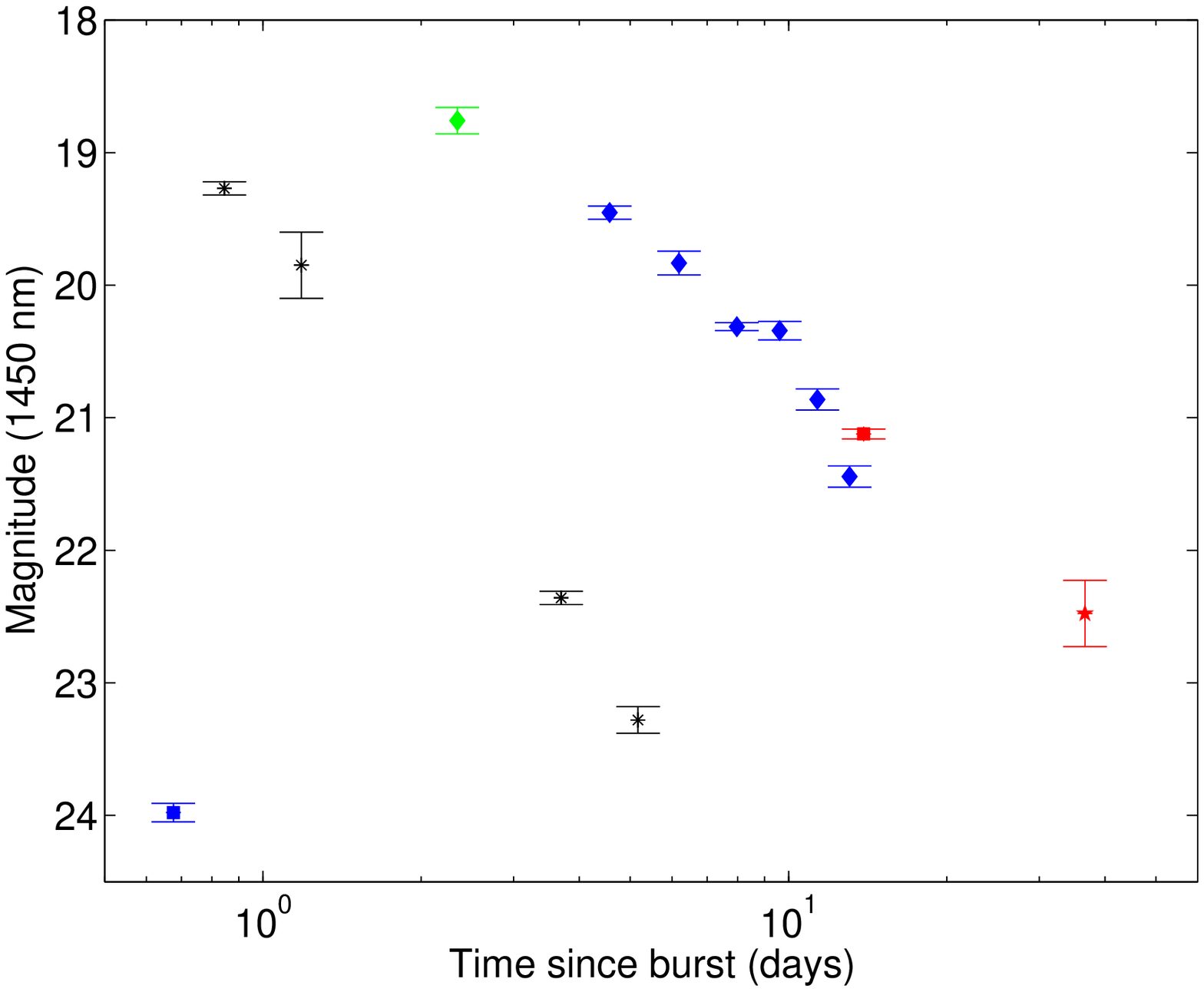}
      \caption{Light curves corrected for distance and cosmological effects. We use the same symbols and colors as in Fig. \ref{Fig_raw}. They are plotted at three different wavelengths : 9000~\AA~(left), 11500~\AA~(middle), and 14500~\AA~(right). See the electronic version for colors.
              }
         \label{Fig_traite}
   \end{figure*}

We selected our sample from all long GRB afterglows observed from Earth in the near infrared (J, H, K bands) between 1997 and the first of January 2008. As pointed out by \citet{klo08}, some discrepancies may occur between publications of data in GCN circulars \citep{bar98} and in refereed journals. Because the latter are more reliable, we considered only data extracted from refereed journals and discarded data from other works. The advantage of near infrared (compared to optical wavelengths) is that extinction is less important \citep{pei92}. We thus reduced the sample to bursts in which the observed afterglow emission was produced (in the burst frame) in the near infrared. This is easily done by computing the emission band using the observation band and the burst redshift. However this last constraint implies a selection bias based on the redshift (see discussion). This reduced sample is listed in Table \ref{table_sample}.

This sample of light curves was then corrected for the extinction due to the Milky Way. We used the work of \citet{she98} for this purpose. Because the extinction due to the host is more uncertain \citep[see e.g.][]{li_08} we have not corrected the host extinction.% (see the section \ref{sec_discu} for the impact of this on the results).}

Finally, the sample was distributed into three bands, based on the rest frame emission wavelength: emission within the $8000-10000$~\AA~range, the $10000-13000$~\AA~ range, and the $13000-16000$~\AA~range. These bands do not correspond to any photometric filter used for observations. We defined them so that they are broad enough to include several afterglows within each band. We used three bands instead of a single large one because this reduces the flux interpolation needed to obtain a final value at a single wavelength (see next section). In the case of \object{GRB 030528}, we corrected the flux for the presence of a bright host galaxy \citep{rau04}. Note however that the host galaxy luminosity is not strongly constrained, thus the error bars on the corrected flux are large. Lastly, we removed from all data the initial part of the afterglow (see Sec. \ref{sec_discu}). These light curves without further corrections are presented in Fig. \ref{Fig_raw}.

\subsection{Distance corrections}

We used the method presented in \citet{gen05} to take into account distance and cosmological effects. This method rescales all light curves to a common redshift of $z = 1$, and thus does not require knowledge of the value of H$_0$. We used a standard flat universe with $\Omega_\Lambda = 0.73$ for the distance correction. Since we already took into account the energy shift due to the cosmological effects by computing the emission band, we we do not need to rescale the observed frequency to a theoretical frequency \citep[as was done previously in the X-rays, and as e.g.][did in the R band]{nar06}. We computed the monochromatic flux at the central frequency of each band, normalizing the fluxes by taking into account the spectral shape of the afterglow (using published spectral indexes or a value of 1.0 in the case of an unknown value). The corrected light curves are presented in Fig. \ref{Fig_traite}.

\section{Results and discussion}
\label{sec_discu}

\subsection{Results}

As one can clearly see in Fig. \ref{Fig_traite}, we observe a hint of clustering of luminosity into two groups. 
From the sample of 19 events, 14 belong to a group of bright events, 3 to a group of dim events, and 2 are isolated. In the following, we will use the notation of \citet{gen08} : bright events in X-ray ({\it x}) or optical ({\it o}) are labeled with a {\it I}, and dim events are labeled with a {\it II}. A burst bright in the optical and dim in X-rays may thus be labeled as {\it xII-oI}. Outliers are labeled with a {\it III}. We note that the infrared groups are identical to the ones observed by \citet{kan06, lia06, nar06} : we will thus not introduce the label {\it ir} but use the label {\it o} to discuss the infrared luminosity clustering. This reinforce the hypothesis that the clustering of infrared luminosity is not spurious. The flux difference between the two group is $\approx$ 3 magnitudes one day after the burst. We observe that \object{GRB 031203}, which does not cluster with the other bursts in the X-ray, seems also not to cluster in the near infrared data (see Fig. \ref{Fig_traite}, right panel). \object{GRB 060614}, another nearby and peculiar burst \citep[see e.g.][]{man07} also does not cluster in luminosity. We discuss these results in the following sections.

\subsection{Selection effects}

As pointed out by \citet{gen08}, the clustering is only apparent in the afterglow data, and the $T_a$ value of \citet{wil07} {\it can be used} to discriminate between early and late light curves. However, most of the afterglows of our sample were observed before the launch of {\em Swift}, and the $T_a$ value is unknown. To be conservative, and at the expense of the sample size, we removed the data taken before 30 000 s after the trigger. This value is larger than most of the $T_a$ values listed in \citet{wil07}. To avoid possible bias near this temporal cut-off, we will discuss in the following values obtained one day after the burst.

The use of three bands instead of a single one is motivated by the reduction of uncertainties resulting from flux interpolations when correcting our sample : an error of $\pm 0.1$ in the spectral index leads to less than 1.2\% error in flux (compared to 5\% in the case of a single large band). This has no incidence on the observed clustering, but reduces the sample size within each band.

A last and most critical selection effect can be seen in Table \ref{table_sample} : we selected only nearby events (compared to the mean {\it Swift} redshift). The bright events group {\it (oI)} has a mean redshift of 1.1. Most of our sample is composed of BeppoSAX events, it is thus not surprising that this low mean redshift is comparable to the BeppoSAX result. However, {\em Swift} has shown that the true redshift distribution of GRBs was different \citep{jac06}. This redshift bias is due to our light curve selection. Removing all events emitted below the I band (but observed in infrared) implies that we remove all events above a redshift of 2.5. Thus, we cannot retain the majority of {\em Swift} bursts within our sample. However, this should not affect our results strongly. As discussed by \citet{gen08}, there are no strong differences in the afterglow properties of very distant and less distant bursts : this bias does not impact our results. However, a possible way to solve this bias is to add further observations within the far infrared bands.

\subsection{Expectations from the model}

The difference between the fluxes of group {\it I} and {\it II} is $\approx$ 3 magnitudes one day after the burst; the optical and infrared groups are identical. This makes us confident that this clustering in luminosity is not spurious despite our small sample. In the following, we will use the results for the optical luminosity clustering given by \citet{nar06}. These authors reported a flux ratio between group {\it I} and group {\it II} of $\sim$ 26. Assuming no spectral variation between optical and near infrared wavelengths, this translate into a difference in magnitude of 3.2, consistent with the observed separation. This implies that the clustering continues towards lower wavelengths than optical bands. 

Previous works have shown from spectral energy distributions of GRB afterglows that optical and near-infrared data can be fit using single power laws \citep[see e.g.][for one example of such works]{sta08}, or in other terms, that the injection frequency $\nu_m$ is lower compared to the infrared bands. We thus have :

\begin{equation}
\label{eq_un}
\nu_m < \nu_{IR}
\end{equation}

with $\nu_{IR} \sim 2 \times 10^{14}$ Hz. In \citet{gen08}, we expressed several constraints on the fireball parameters. Using these constraints and Equation \ref{eq_un}, we can express a constraint on the total energy of the fireball (the constants C are listed in Table \ref{table_constante}) :

\begin{equation}
\label{eq_deux}
\begin{array}{lr}
E_{53} > C n^{-1/5}   \nu_{IR}^{-18/25} & (ISM)\\
E_{53} > C A_*^{-0.5} \nu_{IR}^{-9/10} & (wind).\\
\end{array}
\end{equation}

\begin{table}
\begin{minipage}[t]{\columnwidth}
\caption{Constants indicated in equations \ref{eq_deux} and \ref{eq_trois} for the three classes of afterglows in the wind and ISM.}
\label{table_constante}
\centering
\renewcommand{\footnoterule}{}  % to avoid a line before footnotes
\begin{tabular}{ccc|cc}
\hline \hline
group  & \multicolumn{2}{c}{ISM} & \multicolumn{2}{c}{Wind} \\
       &Eq. \ref{eq_deux} & Eq. \ref{eq_trois} &Eq. \ref{eq_deux}   & Eq. \ref{eq_trois} \\
\hline
xI-oI   & $2.2 \times 10^{8}$ & $4.0 \times 10^{11}$ & $1.2 \times 10^{9}$ & $3.8 \times 10^{12}$\\
xII-oI  & $1.3 \times 10^{7}$ & $2.2 \times 10^{10}$ & $4.5 \times 10^{7}$ & $8.4 \times 10^{10}$\\
xII-oII & $3.4 \times 10^{6}$ & $3.3 \times 10^9$    & $1.4 \times 10^{7}$ & $1.6 \times 10^{10}$\\
\hline
\end{tabular}
\end{minipage}
\end{table}

Clearly, the constraint indicated in Eq. \ref{eq_deux} is not relevant in the case of normal GRBs : even in the case of a wind environment with a very low density (says $A_* = 10^{-4}$) for a burst of the group xI-oI, we have $1.6 \times 10^{-2} < E_{53}$. As one can clearly see in Eq. \ref{eq_deux}, in order to set a better constraint on this value, we need to observe at frequencies of the order of $10^{12}$ Hz, but still above the injection frequency.

\subsection{Impact on the distance estimation}

In \citet{gen08}, we proposed a method to estimate the distance based on the afterglow flux observed one day after the burst in the $2-10$ keV band. However, this method has a degeneracy, because of the existence of several groups. This degeneracy cannot be resolved by optical observations. To clear this problem, we might have used infrared information. However, the fact that we observe exactly the same groups in optical and near infrared makes it clear that we should use longer wavelengths:

\begin{equation}
\label{eq_trois}
\begin{array}{lr}
\nu_m = C E_{53}^{-25/18} n^{-5/18}~Hz & (ISM)\\
\nu_m = C E_{53}^{-10/9} A_*^{-5/9}~Hz & (wind).\\
\end{array}
\end{equation}

We express in Eq. \ref{eq_trois} the position of $\nu_m$ \citep{sar98} for the different groups (the constants C are listed in Table \ref{table_constante}). As one can see, the position of this frequency is different for each group. This implies that at a frequency below the $\nu_m$ value of group xII-oI (the lowest one), we should observe three distinct groups. Observations at such frequencies could be performed by the IRAM observatory or the future ALMA observatory.

\section{Conclusions}

We investigated the clustering in luminosity of Gamma-Ray Burst afterglows using near infrared (J, H, K band) observations of events observed between 1997 and the end of 2007. We corrected the gamma-ray burst afterglow light curve for distance effects and time dilation, and rescaled all light curves to a common distance of $z=1$. We only used observations emitted in the near infrared in the burst frame. We observed a clustering of luminosities and a flux ratio identical to the ones observed in the optical. This implies that the injection frequency is located below the near infrared bands. We confirmed previous works on the clustering of afterglow light curves.

Such a clustering has no clear theoretical explanation yet. In order to explain it, we need to set a strong constraint on the two remaining parameters of the fireball, namely the total energy and the density of the surrounding medium. For the latter, radio observations can be used to impose constraints \citep[see e.g.][]{fra06}. For the former, observations from about 100 GhZ to $10^{13}$ Hz could help. However, this observational window has not been widely used for afterglow observations. It may be interesting to obtain several observations at microwave and millimeter wavelengths that can be coupled with {\em Swift} and robotic telescopes to perform broadband spetro-temporal fitting of the afterglow.

Finally, the constraints on the fireball parameters may be related to the progenitor properties or to other conditions (such as the central engine properties) that would imply a clustering. In order to check the influence of the progenitor, one may add short GRBs (whose progenitor is thought to be a compact object binary) to the current study, that is based only on long GRBs (strongly suspected to be caused by massive stars). A clustering within the same groups would support a physical explanation not related to the progenitor but rather to the central engine of the burst.

\begin{acknowledgements}
BG acknowledges a postdoctoral grant from the French CNES. We would like to thank the anonymous referee for a very constructive report that improved the quality of the paper, and Roger Malina for a careful reading of a draft of this Letter.
\end{acknowledgements}

\end{document}